\documentclass[a4paper, 10 pt, conference]{article}  
\usepackage{hyperref}
\usepackage{graphicx}
\usepackage{amsfonts}
\usepackage{amsmath}
\usepackage[table,xcdraw]{xcolor}
\usepackage{graphicx}

\usepackage{amsmath}
\usepackage{array}
\usepackage{caption}
\usepackage{floatrow}   
\usepackage{subcaption}
\usepackage{graphicx,xcolor} 
\usepackage{breqn}
\usepackage[framemethod=tikz]{mdframed}
\usepackage[tableposition=top]{caption}
\usepackage{float}
\floatstyle{plaintop}
\restylefloat{table}
\usepackage{arydshln}
\usepackage{hyperref}
\usepackage{tabularx}
\usepackage{ifpdf}
\usepackage{booktabs}
\usepackage{booktabs}
\usepackage{graphicx}
\usepackage{adjustbox}
\usepackage{cite}
\usepackage{csquotes}

\title{\LARGE \bf
	Deep Estimation of Speckle Statistics Parametric Images
}

\author{Ali K. Z. Tehrani$^{1}$, Ivan M. Rosado-Mendez$^{2}$, and Hassan Rivaz$^{1}$
	\thanks{This work is supported by the Natural Sciences and Engineering Research Council of Canada (NSERC) RGPIN-2020-04612.}
	\thanks{$^{1}$Ali K. Z. Tehrani and Hassan Rivaz are with the Department of Electrical and Computer Engineering, Concordia University, Montreal, Canada.
		{\tt\small  A\_Kafaei@encs.concordia.ca and hrivaz@ece.concordia.ca}}%
	\thanks{$^{2}$Ivan M. Rosado-Mendez is with the Department of Medical Physics, University of Wisconsin, Madison, USA
		{\tt\small rosadomendez@wisc.edu}}%
}

\begin{document}

	\maketitle
	\thispagestyle{empty}
	\pagestyle{empty}

	\begin{abstract}
		
		Quantitative Ultrasound (QUS) provides important information about the tissue properties. QUS parametric image can be formed by dividing the envelope data into small overlapping patches and computing different speckle statistics such as parameters of the Nakagami and Homodyned K-distributions (HK-distribution). The calculated QUS parametric images can be erroneous since only a few independent samples are available inside the patches. Another challenge is that the envelope samples inside the patch are assumed to come from the same distribution, an assumption that is often violated given that the tissue is usually not homogenous. In this paper, we propose a method based on Convolutional Neural Networks (CNN) to estimate QUS parametric images without patching. We construct a large dataset sampled from the HK-distribution, having regions with random shapes and QUS parameter values. We then use a well-known network to estimate QUS parameters in a multi-task learning fashion. Our results confirm that the proposed method is able to reduce errors and improve border definition in QUS parametric images. 
		\newline
		
	\end{abstract}

	\section{Introduction}
	UltraSound (US) is an imaging modality that has been used widely for clinical applications. The particles that are smaller than the US wavelength scatter US waves. These particles, called scatterers, provide important information about the tissue. Quantitative UltraSound (QUS) studies the acoustic properties of the scatterer by analyzing the backscattererd envelope data \cite{wagner1983statistics,oelze2016review}. 
	
	Spectral and speckle statistics are the two main categories of QUS. In spectral methods, parameters such as backscatter and attenuation coefficients are estimated usually by using a reference phantom \cite{Vajihi2018}. In methods based on speckle statistics, the envelope of Radio Frequency (RF) data is used to obtain different parameters such as scatterer number density and coherent-to-diffuse scattering ratio \cite{wagner1983statistics,Rosado2016}. These parameters are quantified by modeling the first order statistics of the envelope data using different probability distribution functions.
	The Scatterer number density, which is defined as the number of scatterers inside the resolution cell (the beam profile -6 dB point), is an important property that affects the speckle statistics. 
	
	If there are many (more than 10) randomly distributed scatterers inside the resolution cell, the envelope data can be described as fully developed speckle (FDS) and otherwise as under developed speckle (UDS) \cite{wagner1983statistics}. The Rayleigh Probability density function can be used to model FDS but it fails when the envelope data is UDS or structured (producing coherent scattering) \cite{wagner1983statistics,Rosado2016}. The Homodyned K (HK)-distribution is able to model different scattering scenarios (including low and high number of scatterers as well as coherent scattering). This distribution is complex and does not have a closed-form expression \cite{destrempes2013estimation,hruska2009improved}. The Nakagami distribution has also been widely used to model different scattering conditions due to its simplicity and having a closed-form expression \cite{shankar2001classification,tsui2017effect}. 
	
\begin{figure} 
		\centering
		\includegraphics[width=0.7\textwidth]{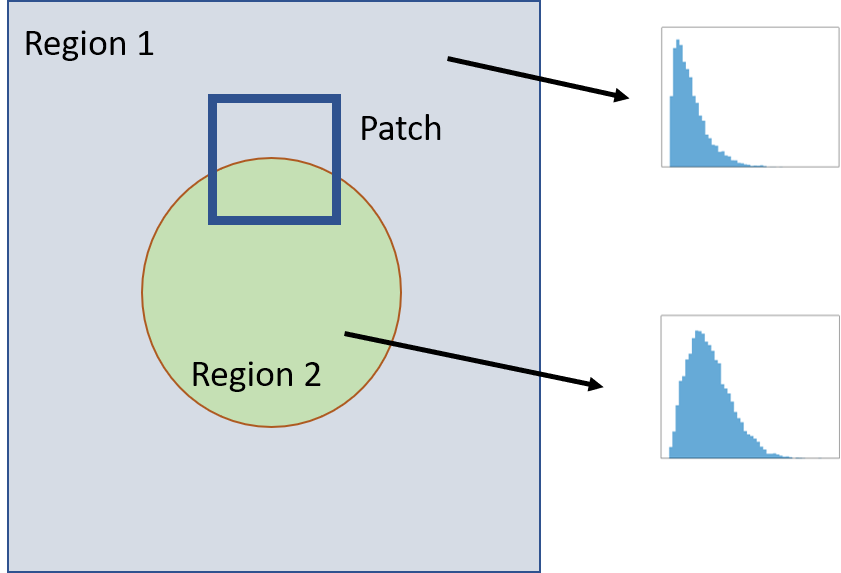}
		\caption{Different regions and their corresponding histograms are specified. The patch located on the boundary contains two regions with different distributions.}
		\label{fig:boundary}
	\end{figure}

	QUS parametric images provide information on the spatial variability of different QUS parameters throughout the US image and provide additional information to B-mode imaging. QUS parametric images can be formed by dividing the US envelope data into small overlapping patches and calculating the QUS parameters such as Nakagami and HK-distributions parameters of those patches. Accurate estimation of HK-distribution requires a large amount of samples. Consequently, large patches should be employed which results in low spatial resolution. Another challenge in QUS parametric image estimation is that the data inside each patch is assumed to come from one distribution. This is not true especially on the boundaries of different tissue types, resulting in unreliable values within those patches. This problem is illustrated in Fig. \ref{fig:boundary}.

	Convolutional Neural Networks (CNN) have been successfully employed for QUS \cite{zhang2020deep,tehrani2021ultrasound}. In a recent work, CNNs were employed to classify patches of US envelope data into FDS and UDS \cite{tehrani2021ultrasound}, which can suffer from inhomogeneity within each patch. It should be noted that CNNs also have a receptive field, usually a large region of the image that is utilized for inferring the properties of the center pixel. The difference between the receptive field and a patch is that complex weights are trained for different parts of the receptive field, giving the network an ability to theoretically ``see’’ different distributions within a receptive field \cite{behboodi2020receptive}.
	
	In this work, we aim to estimate QUS parametric images of parameters from the HK and the Nakagami distributions using a multi-task learning (MTL) method \cite{zhang2018overview}. Our hypothesis is that by using CNNs, we can mitigate the aforementioned problems associated to the use of patches in the reconstruction of QUS parametric images which leads to higher spatial resolution and more accurate parameter estimation around the boundaries.
	\vspace{-0.1cm}%
	\section{Materials and Methods}
	\vspace{-0.1cm}%
	\subsection{Nakagami and HK-distribution}
	The Nakagami distribution has been widely utilized to model envelope data having different scattering conditions. This distribution can be written as \cite{tsui2017effect}: 
	\begin{equation}
		P_{N} (x| m, \Omega) = \frac{2m^mx^{2m-1}}{\Gamma(m)\Omega^m}exp(-\frac{m}{\Omega}x^2)U(x)
	\end{equation}
	where $\Gamma$ and $U(.)$ denote Gamma and step functions, respectively. $x$ is the amplitude, $m$ controls the shape of the distribution and $\Omega$ is the scale parameter. The parameter $m$, which is also called Nakagami parameter, is linearly related to the scatterer number density when the scatterer number density is low \cite{tsui2015artifact} (a low scatterer number density results in $m<1$). When the scatterer number density increases, $m$ approaches the value of 1. When there are coherent scattering, $m$ can be higher than 1. The parameter $\Omega$ is sensitive to the scattering power of the envelope data. These parameters can be obtained using the Maximum Likelihood Estimation (MLE) method \cite{cheng2001maximum}.
	
	The HK-distribution equation can be written as \cite{hruska2009improved,destrempes2013estimation}:
	\begin{equation}
		P_{HK}(x|\varepsilon ,\sigma^2,\alpha) = x\int_{0}^{\infty}uJ_{0}(u\epsilon)J_{0}(ux)(1+\frac{u^2\sigma^2}{2})^{-\alpha}du
	\end{equation}
	where $\alpha$ depends on the scatterer number density and $x$ is the amplitude. The diffuse signal power can be obtained by $2\sigma^2\alpha$ and the coherent signal power is $\epsilon^2$  \cite{destrempes2013estimation}. The estimation of the ratio of coherent to diffuse signal power ($k$) and $\alpha$ has been an active area of research. The XU estimator is considered as a state-of-the-art (SOTA) algorithm for HK-distribution parameter estimation. Please refer to \cite{destrempes2013estimation} for more information about this algorithm.
	
	Both $m$ and $\alpha$ can be employed to describe the scatterer number density \cite{destrempes2013estimation,tsui2017effect}. $m$ requires less data to be estimated compared to $\alpha$ but it saturates when the scatterer number density is close to 10, whereas $\alpha$ can be used to distinguish higher scatterer number densities. In this paper, we aim to reconstruct parametric images of $log_{10}(\alpha)$ (similar to \cite{zhou2021parameter}, $log_{10}$ of $\alpha$ is estimated) and $m$. The proposed method can be used to estimate other QUS parameters such as $\sigma$ and $k$ as well.   
	\vspace{-0.1cm}%
	\subsection{Simulation Data and Experimental Phantoms}
	\vspace{-0.1cm}%
	\subsubsection{Simulation data generation}
	A diverse and large dataset is required to train a network to estimate QUS parametric images without patching. We first generate binary random shapes. These binary masks are employed to place US samples having different distributions. Each sample represents one realization of the echo produced by various scatterer within the resolution cell volume. 
	
	We use the algorithm presented in \cite{hruska2009improved} to sample from HK-distribution:
	\begin{equation}
		a_i = \sqrt{\left (  \varepsilon +X\sigma \sqrt{Z/\alpha }\right )^2+\left (  Y\sigma \sqrt{Z/\alpha }\right )^2}
	\end{equation}
	where $a_i$ is the generated HK-distribution sample, $X$ and $Y$ are independent and identically distributed (i.i.d) samples of unit Normal distribution and $Z$ is sampled from the Gamma distribution with shape parameter $\alpha$ and scale parameter of 1. To generate the simulation data, we assume that the incoherent scattering power ($\sigma$) and $\alpha$, which is highly related to scatterer number density, can vary while $\varepsilon=0$. 
	
	\vspace{-0.01cm}%
	\subsubsection{Homogeneous Phantoms}
	\vspace{-0.01cm}%
	We used two experimental phantoms having different scatterer number densities and different scattering cross section. The phantoms had the size of 15cm$\times$ 5cm$\times$ 15cm and were built using homogeneous mixture of agarose gel media and glass beads as scattering agents. The phantom B (Medium) and C (Low) had 9 and 3 scatterers per $mm^3$ with dimension range of 75-90 and 126-151 $\mu m$, respectively. The phantoms are named based on \cite{tehrani2021ultrasound} which proposed the patch-based classification of the scatterer number density of these phantoms.
	
	The phantoms were imaged using an Acuson S2000 scanner (Siemens Medical Solutions, Malvern, PA) with a 18L6 transducer operating at 8.89 MHz center frequency and sampling frequency of 40 MHZ.


	\subsection{Deep Estimation Method}
	The proposed method can be described as the following. First, 15,000 (14,000 for training and 1000 for validation) US images having independent samples drawn from HK-distribution were generated. Furthermore, we train a SOTA network using the generated data to predict $log_{10}(\alpha)$ and $m$. In the last step, the simulation and experimental data were evaluated using the trained network. We employed the DeeplabV3 network \cite{yurtkulu2019semantic} which has shown promising results in semantic segmentation and used an NVIDIA TITAN V with 12 GB of RAM to train the network. The following loss function is utilized to train the network:

	\begin{equation}
		loss = (\left |  log_{10}(\widetilde{\alpha}), log_{10}(\alpha)\right |)^{1/2}+ \lambda \left \{\left |  \widetilde{m}, m\right |  \right \} 	
	\end{equation}
	where $|.|$ and $\widetilde{(.)}$ denote smooth L1 norm and the predicted value, respectively. $log_{10}(\alpha)$ and $m$ are the ground truth values of log compressed $\alpha$ and Nakagami parameter. The power 1/2 is employed for smooth L1 loss of $log_{10}(\alpha)$ to penalize small differences between the ground truth and predicted value. We emprically set $\lambda=0.1$ to give more importance to estimation of $log_{10}(\alpha)$ since it is a more challenging parameter to be estimated compared to $m$.
	
	We compare our method with patch-based estimation of QUS parametric images. To be specific, the SOTA XU estimator \cite{destrempes2013estimation} is used to estimate $\alpha$ and MLE to estimate $m$ \cite{cheng2001maximum}. The size of generated independent samples for each image is $256\times128$ and a patch around the sample of interest ($32\times16$) is extracted to obtain QUS parameters for the patch-based methods. 
	Relative Root Mean Square Error (RRMSE) of the estimated parameters is employed as the metric for the comparison. RRMSE can be defined as \cite{zhou2021parameter}:
	\begin{equation}
		RRMSE(\phi,\widetilde{\phi})=\frac{\sqrt{\left \langle (\widetilde{\phi}-\phi)^2 \right \rangle}}{|\phi|}
	\end{equation}
	where $<.>$ denotes averaging operation and $\widetilde{\phi}$ is the predicted parameter.


	\vspace{-0.1cm}%
	\section{Results}
	\vspace{-0.1cm}
	\subsection{Simulation Results}
	\vspace{-0.1cm}
	\begin{figure}[t]
		\centering
		\includegraphics[width=0.9\textwidth]{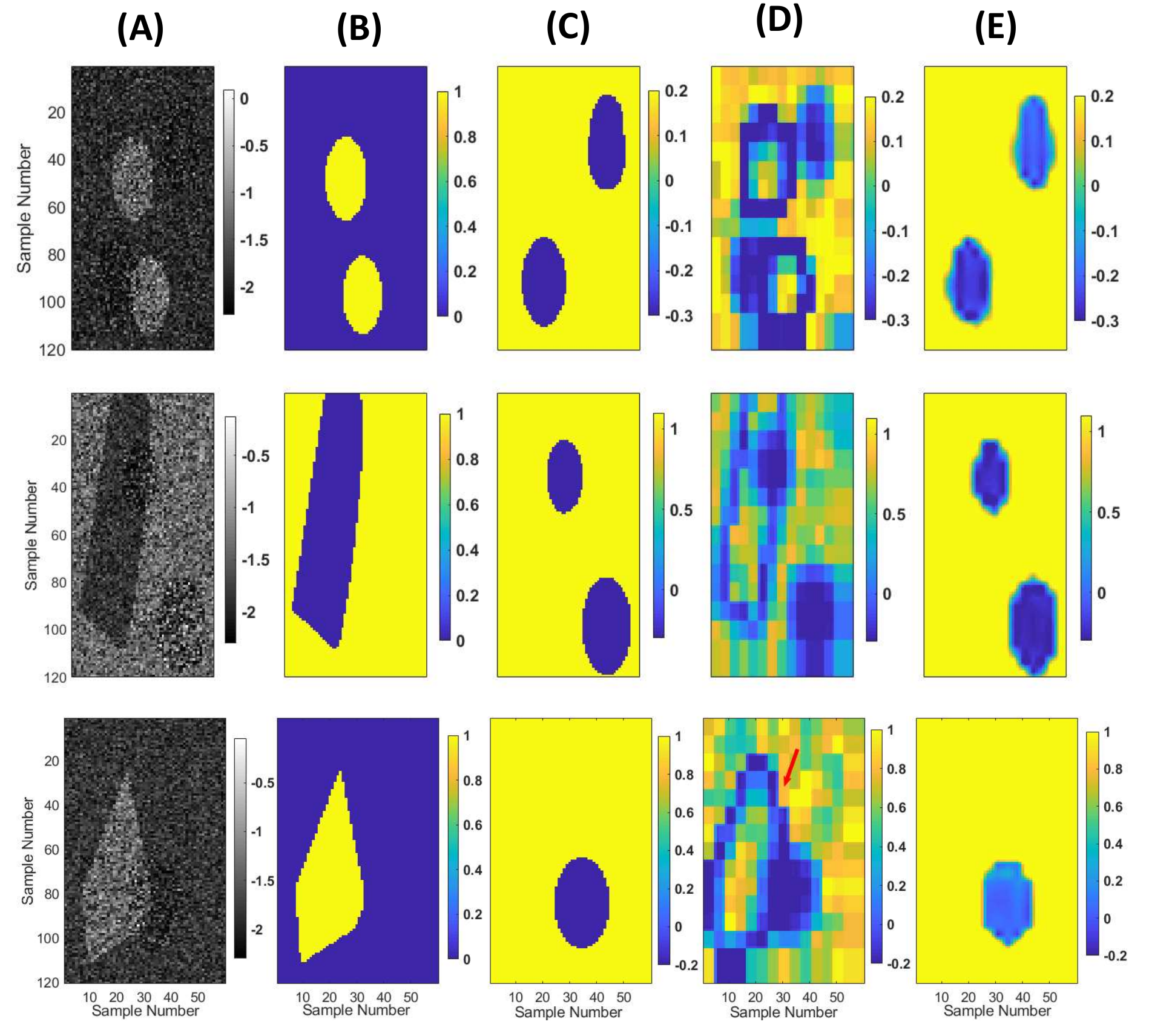}
		\caption{Parametric images of $log_{10}(\alpha)$ with different estimators. B-mode images (A), $\sigma$ (B), $log_{10}(\alpha)$ (C), reconstructed $log_{10}(\alpha)$ using patch-based XU method (D) and deep estimation (E). The XU estimators (patch size: $3mm \times 3 mm$) does not provide reliable results on the boundaries having different distributions. When $\sigma$ changes but $\alpha$ remains constant, XU estimator predicts variations on $\alpha$ (marked with arrows). Whereas, the proposed method provides reliable estimation of $log_{10}(\alpha)$. }
		\label{figure:sim_res}
	\end{figure}

	The mean and standard deviation of RRMSE are reported for a test set with 200 samples in Table \ref{tab:sim}. We exclude the regions where $log(\alpha)=0$ similar to \cite{hruska2009improved,zhou2021parameter}.  
	According to the Table, the proposed method has substantially lower error compared to the patch-based estimation (the improvements with respect to the patch-based method are given in the last row). The results demonstrate the superiority of the proposed method when there are a few independent samples.
	
\begin{figure}[t]
		\centering
		\includegraphics[width=0.7\textwidth]{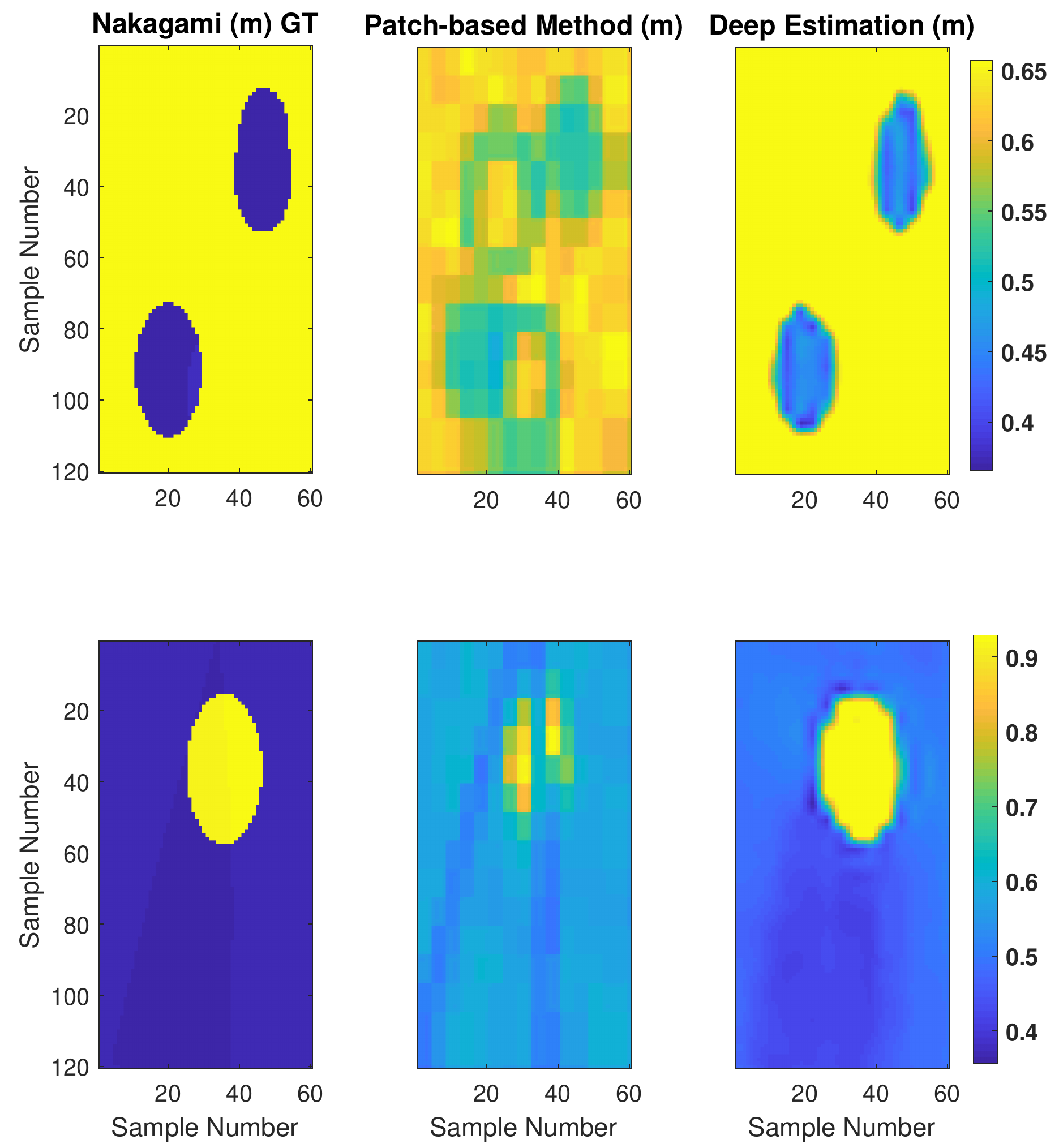}
		\vspace{-0.1cm}%
		\caption{Two reconstructed Nakagami parametric images of the test data; The ground truth (left), patch-based estimation of parametric images (middle) and the deep estimation method (right). }
		\label{figure:sim_res_fe}
	\end{figure}

	
	A few examples of reconstructed parametric images of $log_{10}(\alpha)$ are shown in Fig. \ref{figure:sim_res}. It can be seen that at the boundaries between regions with different scattering conditions, the patch-based method results in unreliable values, while the proposed method is able to provide clearly-defined borders. Two examples of estimated parameters of Nakagami distribution are illustrated in Fig. \ref{figure:sim_res_fe}. Patch-based estimation of the Nakagami parameter ($m$) has some variations in regions where the ground truth $m$ is fixed. The reason is that in those regions, diffuse scattering power ($\sigma$) was varied, hence the distribution changed. The same effect can be seen in Fig. \ref{figure:sim_res} where incorrect values of $log_{10}(\alpha)$ are obtained by the XU estimator due to the boundary problem. The deep estimator does not suffer from the aforementioned problem and produces reliable QUS parametric images.

	\vspace{-0.05cm}
	\begin{table}[]
		\caption{mean and standard deviation of RRMSE of test simulation data.}
		\label{tab:sim}
		\resizebox{0.85\textwidth}{!}{%
			\begin{tabular}{@{}ccc@{}}
				\toprule
				& \textbf{$log_{10}(\alpha)$}  & \textbf{$m$}         \\ \midrule
				Patch-based Estimation     & 0.340$\pm$0.646 & 0.145$\pm$0.072 \\
				Deep Estimation         & 0.131$\pm$0.356 & 0.0863$\pm$0.054 \\ \midrule
				Improvement (\%) & 61.4            & 40.5                       
		\end{tabular}}
	\end{table}
	\vspace{-0.05cm}
	\vspace{-0.05cm}
	\vspace{-0.010cm}
	
	\begin{figure}[]
		\centering
		\includegraphics[width=0.99\textwidth]{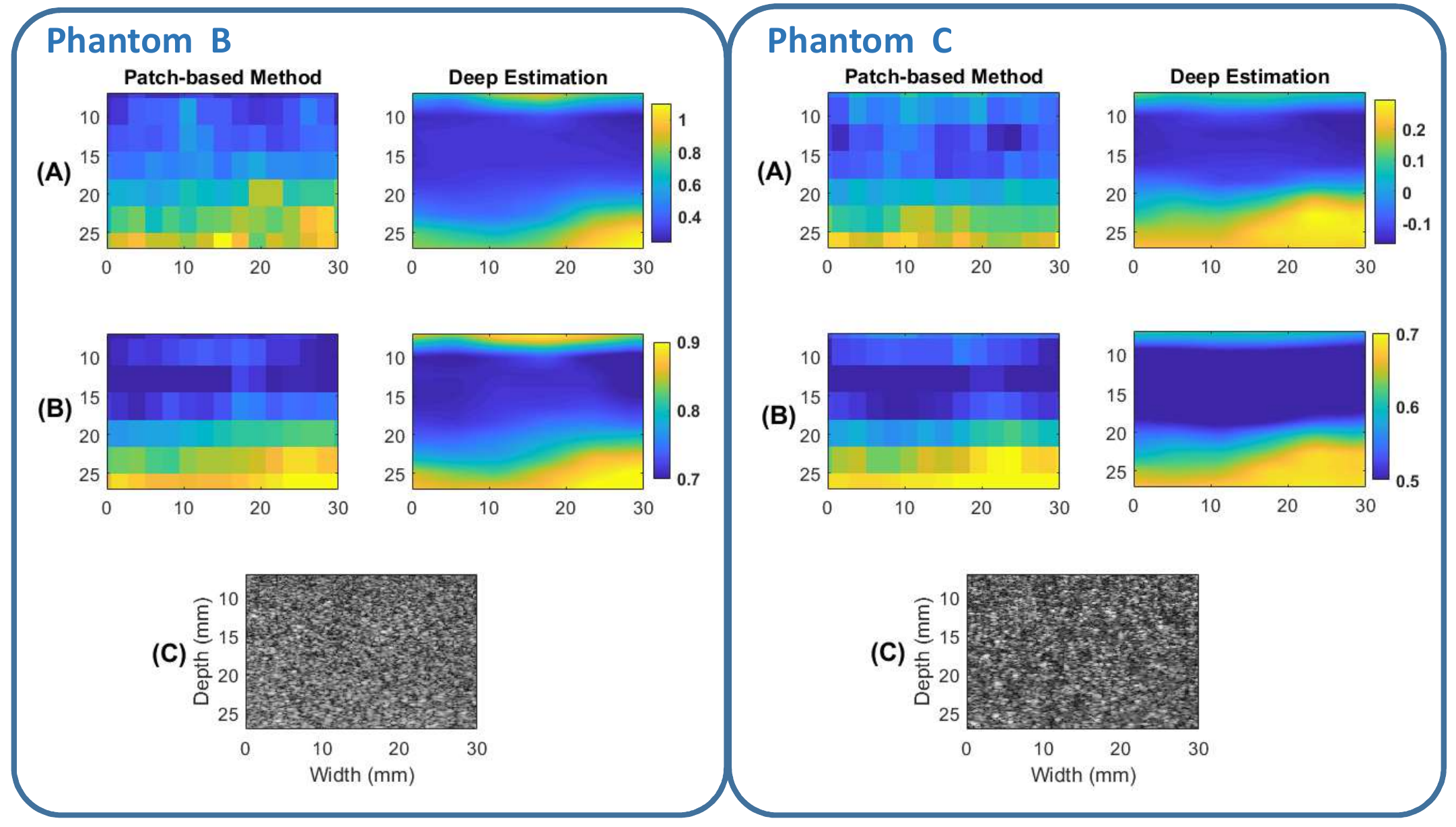}
		\vspace{-0.2cm}%
		\caption{The phantoms B and C QUS parametric images. $log_{10}(\alpha)$ (A), Nakagami $m$ (B) and the B-mode image (C). The first column denotes the results of conventional patch-based methods and the second column denotes the results of the deep estimation method.}
		\label{figure:exp_medium}
	\end{figure}
	\vspace{-0.02cm}
	\subsection{Homogeneous Phantoms Results}
	\vspace{-0.02cm}
	The QUS parametric images of the phantom B and C are shown in Fig. \ref{figure:exp_medium}. The results of both patch-based and deep estimation methods are averaged across 18 frames. While the results of the proposed method agree with the patch-based estimation, our proposed algorithm provides higher resolution QUS parametric images compared to the patch-based extraction of QUS parametric image.

	The exact value of $\alpha$ is unknown in these phantoms since the resolution cell varies with depth. We expect to have low values of $\alpha$ and $m$ at the focal point and higher values at the bottom (due to beam divergence). The $log_{10}(\alpha)$ of phantom B at the focal point is around 0.3 using our network,  which increases to 0.8 at the bottom. The same effect can be seen for the phantom C, the $log_{10}(\alpha)$ is increased from -0.145 at the focal point to 0.23 at the bottom. This increase can be associated to the beam divergence at the bottom which results in the expansion of the resolution cell. 
	
	
	In experimental phantom results, the exact values of the reconstructed QUS parametric images are not known; therefore, to investigate the accuracy of the reconstructed QUS parametric images, we perform the following procedure. The parameters $m$ and $\alpha$ are highly related to each other; therefore, a very high correlation between these parameters is expected when data is not FDS ($m$ is not saturated). However, the noise caused by estimation error reduce this correlation. The correlation between these parameters for the two phantoms are given in Table \ref{tab:corr}. It can be seen that the correlation coefficients are higher for the QUS parametric images reconstructed by the deep estimation method than the patch-based ones especially for phantom C which has lower scatterer number density.

	\vspace{-0.025cm}%
	\begin{table}[]
		\resizebox{0.8\textwidth}{!}{
			\caption{Correlation between $\alpha$ and $m$ for the two experimental phantoms.}
			\label{tab:corr}
			\begin{tabular}{@{}ccc@{}}
				\toprule
				Parameter              & Phantom B      & Phantom C      \\ \midrule
				Patch-based Estimation & 0.919          & 0.922          \\
				Deep Estimation        & \textbf{0.920} & \textbf{0.986} \\ \bottomrule
		\end{tabular}}
	\end{table}
	\vspace{-0.025cm}%
	\section{Conclusion}
	\vspace{-0.025cm}%
	In this paper, we provide a general framework based on CNN to estimate QUS parametric images without patching. We generate US envelope data sampled from HK-distribution. These samples are composed of regions having different values of scatterer number density and diffuse scattering power. We train a CNN using the generated data to estimate the parameter $log_{10}(\alpha)$ and $m$.  Our results confirm that the proposed method can obtain accurate estimation of QUS parametric images and outperforms patch-based estimators.

	\vspace{-0.05cm}%
	\section{Acknowledgment}
	\vspace{-0.05cm}%
	We acknowledge the support of the Natural Sciences and Engineering Research Council of Canada (NSERC) RGPIN-2020-04612.
	\vspace{-0.05cm}
	
	\vspace{-0.05cm}
	\bibliographystyle{IEEEtran}
	\bibliography{refs3}

\end{document}